\documentclass[twocolumn,aps,prl,showpacs,groupedaddress,floatfix,letterpaper]{revtex4}
\pdfoutput=1
\usepackage{ae}
\usepackage{graphicx}
\usepackage{hyperref}
\usepackage[leqno]{amsmath}
\begin{document}
\title{Trapped-ion quantum logic gates based on oscillating magnetic fields}

\author{C. Ospelkaus}
\author{C. E. Langer}
\altaffiliation[Present address: ]{Lockheed Martin Space Systems Company, Littleton, CO}
\author{J. M. Amini}
\author{K. R. Brown}
\author{D. Leibfried}
\author{D. J. Wineland}
\affiliation{National Institute of Standards and Technology; 325 Broadway, Boulder, Colorado 80305, USA}

\begin{abstract}
Oscillating magnetic fields and field gradients can be used to implement single-qubit rotations and entangling multi-qubit quantum gates for trapped-ion quantum information processing (QIP). With fields generated by currents in microfabricated surface-electrode traps, it should be possible to achieve gate speeds that are comparable to those of optically induced gates for realistic distances between the ion crystal and the electrode surface. Magnetic-field-mediated gates have the potential to significantly reduce the overhead in laser beam control and motional state initialization compared to current QIP experiments with trapped ions and will eliminate spontaneous scattering, a fundamental source of decoherence in laser-mediated gates. 
\end{abstract}


\maketitle

Most current schemes for quantum information processing with trapped ions use laser-induced interactions to implement single-qubit rotations and entangling multi-qubit gates by coupling the internal states of ions to their motion~\cite{Cirac1995a}. While these schemes have been successful in manipulating small numbers of ions, scaling up to larger register sizes and/or multiple quantum registers will require a large overhead in laser-beam power and control. Moreover, operation fidelities will eventually determine the necessary overhead for quantum error correction, making it desirable to minimize gate imperfections. For laser-induced gates, fidelities are currently limited by fluctuations of classical parameters, such as noise in laser beam intensity at the ion positions. For gates induced by stimulated-Raman transitions, spontaneous emission poses an additional fundamental limit~\cite{Ozeri2007a}.

One possible approach to mitigate the overhead of laser beam control replaces the precise timing of laser beams with the precise control of ion transport through stationary beams. This could also reduce power requirements by utilizing the same laser beam for parallel operations~\cite{Leibfried2007a}. An alternative to laser-induced gates would be to use magnetic fields and their gradients to perform gate operations, thereby eliminating spontaneous-emission decoherence during operations. As discussed in~\cite{Mintert2001a,Johanning2007a}, applying a static magnetic field gradient provides qubit addressing and coupling of motional and internal quantum states as required for two-ion entangling gates. Reference~\cite{Leibfried2007a} also considered implementing multi-qubit gates by transporting ions over a microfabricated static magnetic field pattern. This takes advantage of the small length scales of microfabricated surface-electrode traps to realize strong oscillating magnetic field gradients in the rest frame of the ion, and could thereby achieve gate speeds comparable to those of currently demonstrated laser-mediated gates \cite{Leibfried2003a}. 

Coupling between internal and motional states is particularly strong with laser beams (wavelength $\lambda$) where the length scale for field gradients is given by $\lambda/2\pi$.  However, if current-carrying structures can be made small enough, oscillating magnetic fields can also be used to implement this coupling~\cite{Wineland1998a}, as in the $g-2$ experiments of Dehmelt and co-workers~\cite{Dehmelt1990a}. In the context of coupled-spin Hamiltonian simulation, Ref.~\cite{Chiaverini2007a} discussed the use of circular coils in a surface array to implement these Hamiltonians with oscillating magnetic fields. 

In this letter, we discuss the implementation of single- and multi-qubit logic gates with oscillating magnetic fields in the context of universal computation. We analyze gates based on $\sigma_\varphi \sigma_\varphi$~\cite{Molmer1999a,Sorensen2000a,Solano1999a} and $\sigma_z \sigma_z$~\cite{Leibfried2003a,Milburn2000a} couplings, where $\sigma_i$ are Pauli operators, and $\sigma_\varphi = \cos{\varphi}\,\sigma_x + \sin{\varphi}\,\sigma_y$. We also consider the integration of the field-producing conductors with the requirements for surface-electrode traps ~\cite{Seidelin2006a,Labaziewicz2008a} and compare these gates with those implemented with laser beams. 

We consider a string of $N$ ions of mass $m$, aligned along the $y$ axis. We assume the trap axes orthogonal to $y$ are aligned along the $x$ and $z$ directions. Two internal states $\left|\uparrow\right>$ and $\left|\downarrow\right>$ of each ion compose a qubit with transition frequency $\omega_0$. We choose the $z$ axis as the quantization axis, provided by a static magnetic field $B_0{\vec e}_z$. Let $q_n$ be the displacement (along $x$ or $z$) of ion $n$ relative to its equilibrium position ($q_n=0$). We write $q_n$ in terms of normal modes with coordinates $\tilde{q}_j$ and frequencies $\omega_j$: $q_n=\sum_j{b_{j,n}{\tilde q}_j}$~\cite{James1998a}. With ${\tilde q}_0^j:=\sqrt{\hbar/(2m\omega_j)}$, we have ${\tilde q}_j={\tilde q}_0^j({\hat a}_j + {\hat a}_j^\dagger)$, where ${\hat a}_j^\dagger$ and ${\hat a}_j$ are normal-mode creation and annihilation operators. The interaction-free Hamiltonian for the qubit states and the motion can be written as $H_0=\sum_n{\frac{\hbar\omega_0}{2}\sigma_z^n} + \sum_j{\hbar\omega_j{{\hat a}_j^\dagger}{\hat a}_j}$. An oscillating magnetic field ${\vec B}(q,t)={\vec e}_x B_x + {\vec e}_z B_z = [{\vec e}_x ({\tilde B}_x + q \, {\partial {\tilde B}_x}/{\partial q}\bigr|_{q=0}) + {\vec e}_z ({\tilde B_z} + q\, {\partial {\tilde B}_z}/{\partial q}\bigr|_{q=0})] \cos(\omega t+\varphi)$ couples the internal states and the motion. The interaction-picture magnetic dipole Hamiltonian for ion $n$ in the rotating-wave approximation is given by
\begin{multline}
  \label{eq:hamiltonian}
  H_{I,n} = - \hbar e^{-i(\omega t+\varphi)} \Bigl[ \left( \Omega^x       \sigma_+^n e^{i\omega_0 t} + \Omega^z       \sigma_z^n \right) + {}\\
  \sum_j                                            \left( \Omega_{j,n}^x \sigma_+^n e^{i\omega_0 t} + \Omega_{j,n}^z \sigma_z^n \right) 
                                                    \left( e^{-i\omega_j t}{\hat a}_j + e^{i\omega_j t}{\hat a}_j^\dagger \right)\Bigr] + h. c.,  \\
\end{multline}
where the Rabi frequencies are defined by:
\begin{align*}
  \Omega^x
  &= \frac{{\tilde B}_x|_{q=0}}{2\hbar}\left<\downarrow|\mu_x|\uparrow\right>
  &\Omega_{j,n}^x
  &= \frac{b_{j,n} {\tilde q}_0^j}{2\hbar}\frac{\partial {\tilde B}_x}{\partial q}\Bigr|_{q=0}\left<\downarrow|\mu_x|\uparrow\right>\\
  \Omega^z
  &= \frac{{\tilde B}_z|_{q=0}}{2\hbar}\left<\uparrow|\mu_z|\uparrow\right>
  &\Omega_{j,n}^z 
  &= \frac{b_{j,n} {\tilde q}_0^j}{2\hbar}\frac{\partial {\tilde B}_z}{\partial q}\Bigr|_{q=0}\left<\uparrow|\mu_z|\uparrow\right>
\end{align*}
$\vec\mu$ is the magnetic moment of an ion, and for simplicity, we have assumed $\left<\uparrow|\mu_z|\uparrow\right>=-\left<\downarrow|\mu_z|\downarrow\right>$ here. The $\Omega^z$ term in~(\ref{eq:hamiltonian}) describes periodic changes in the energy of each ion-qubit due to the oscillating $z$-component of the magnetic field. The $\Omega^x$ term describes driven Rabi oscillations applied simultaneously to all ion-qubits (``carrier'' transitions) if the driving field frequency is close to $\omega_0$. Both terms implement global single-qubit rotations. 

The remaining terms in~(\ref{eq:hamiltonian}) describe simultaneous changes of internal and motional states and can therefore be used to implement multi-ion entangling gates. From the $\Omega_{j,n}^x$ term, we obtain spin-flip transitions simultaneous with the creation or annihilation of phonons in motional mode $j$ (for $\omega\approx\omega_0\pm\omega_j$). For a single ion, $\Omega_{j,n}^x$ is the Rabi flopping frequency for the first ``blue and red sideband'' transitions ($\omega=\omega_0\pm\omega_j$). The $\Omega_{j,n}^z$ term, on the other hand, provides qubit-state-dependent excitation of motional mode $j$ without simultaneous spin flips. 

To implement multi-qubit gates, we first consider an oscillating $B_z$ field close to a normal-mode frequency $\omega_j$ ($\omega=\omega_j-\delta$), and keep only near-resonant terms in~(\ref{eq:hamiltonian}). Under the influence of this periodic force, the motional state of the ion follows a circular trajectory in phase space~\cite{Leibfried2003a}. At $\tau=2\pi/\delta$, the motional state has returned to the origin, and the propagator is
\begin{equation}
  \mathcal{U}_{zz}(\tau)=\exp[\frac{2\pi i}{\delta^2}(\sum_n{\Omega_{j,n}^z}\sigma_z^n)^2] \quad .
\end{equation}
For a two-ion ``rocking'' mode  ($b_{j,1} = - b_{j,2}$) and the detuning $\delta/4=\Omega_{j,n}^z$, we obtain a $\sigma_z\sigma_z$ phase gate equivalent to Refs.~\cite{Leibfried2003a,Milburn2000a}. The $\sigma_z\sigma_z$ gate requires the $\left|\uparrow\right>$ and $\left|\downarrow\right>$ states to have a different magnetic moment and can therefore not be directly applied to qubit states that are magnetic-field insensitive to first order, which is desirable for long coherence times~\cite{Langer2005a,Lee2005a}. However, with the use of efficient single-qubit rotations as discussed here, it is possible to temporarily transform from the field-independent qubit-state manifold and perform the gate.

The $\Omega_{j,n}^x$ term can be used to implement the $\sigma_\varphi\sigma_\varphi$ gate~\cite{Molmer1999a,Sorensen2000a,Solano1999a}. We can simultaneously apply two equal-amplitude $B_x$ fields at frequencies detuned from the first blue and red sideband by $\delta$ ($\omega_{b,r}=\omega_0\pm\omega_j\mp\delta$) with phases $\varphi_b$ and $\varphi_r$, and keep only resonant multi-qubit spin-flip terms. Again, at $\tau=2\pi/\delta$, the motional mode has returned to its initial state, and the propagator is 
\begin{equation}
  \mathcal{U}_{\varphi\varphi}(\tau)=\exp[\frac{2\pi i}{\delta^2}(\sum_n{\Omega_{j,n}^x\sigma_{\varphi_s}^n})^2] \quad ,
\end{equation}
where $\varphi_s=(\varphi_b+\varphi_r)/2$. We thus obtain the equivalent propagator to the $\sigma_z\sigma_z$ gate, which now acts on the single-qubit eigenstates of $\sigma_{\varphi_s}^n$.

To estimate the coupling strengths, consider a current $I$ running through a thin straight wire aligned along the positive $y$ axis. This produces a field $(-I\mu_0/(2\pi d)){\vec e}_z$ at $x=d$ and field gradients $\partial B_x/\partial z=\partial B_z/\partial x=I\mu_0/(2\pi d^2)$. For an ion located at this position, Rabi frequencies for single-qubit rotations are therefore proportional to $1/d$, whereas Rabi frequencies for entangling gates scale as ${\tilde q}_0^j/d^2$. As an example, for $^9$Be$^+$ and $\omega_j=2\pi\cdot5$\,MHz, ${\tilde q}_0^j\approx 10$\,nm, which is small compared to current values of ion-to-electrode distances $d$ (25 to 100\,$\mu$m). Therefore, significant off-resonant carrier excitation ($^x\Omega_0$) and ac-Zeeman shifts would occur during multi-qubit gates unless current geometries with a strong gradient and a small absolute value of the field are used. 

\begin{figure}[tb]
  \centering
  \includegraphics[width=\columnwidth]{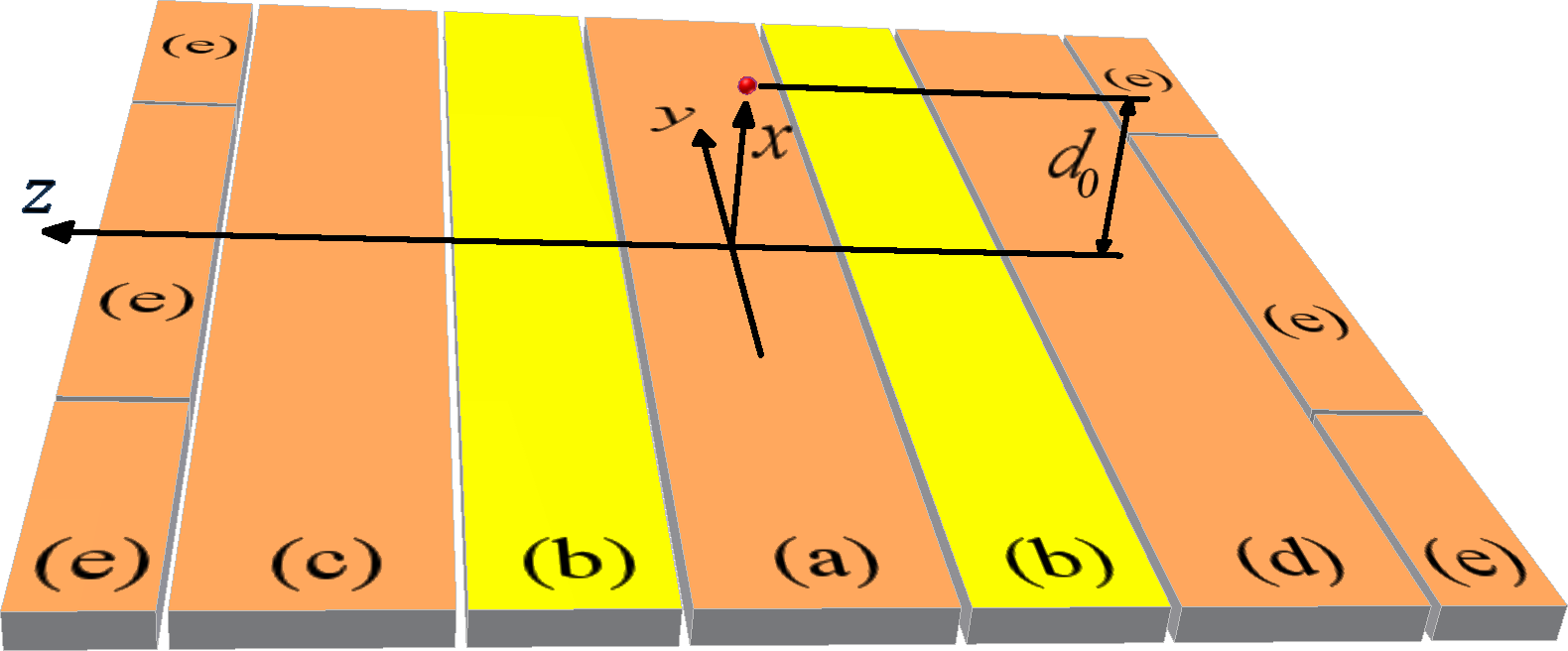}
  \caption{({\bf color online}) Example geometry for driving ion qubit gates with oscillating magnetic fields produced by surface-electrode traps. The trap electrodes lie in the $yz$ plane. The ion is assumed to be trapped at a distance $d_0$ above the trap plane. Oscillating currents in electrodes (a), (c) and (d) can be used to implement single-qubit rotations and entangling gates (see text).}
\label{fig:sample_geometry.png}
\end{figure}

We now examine possible current geometries compatible with surface-electrode traps. For single-qubit rotations, we require oscillating fields transverse to $B_0$. To implement the $\sigma_z\sigma_z$ multi-qubit gate, the ions need to sample changes in $B_z$ during their motion; for the $\sigma_\varphi\sigma_\varphi$ gate, the ions need to sample changes in $B_x$ or $B_y$ during motion. For either gate we desire $B_x = B_z = 0$ at the ions. Fig.~\ref{fig:sample_geometry} shows how these requirements can be integrated in an example surface-electrode trap. The trapping potential of this ``five-wire'' surface trap~\cite{Chiaverini2005a} is provided by a central static potential electrode (a), two rf electrodes (b), additional static potential electrodes (c), (d), and segmented outer electrodes (e) providing confinement along $y$. The ion is trapped at the rf pseudopotential null line at a distance $d_0$ from the surface. The electrode dimensions along $z$ are chosen so that antiparallel currents $I(t)=\tilde I \cos (\omega t + \varphi)$ through electrodes (c) and (d) provide a field $\tilde{B}_x = 1.5\cdot10^{-7} \cdot I(t)/d_0$\,T at the ion to implement rotations, where $I$ and $d_0$ are expressed in amperes and meters, respectively. For $d_0=30\,\mu$m and using the field-independent $^9$Be$^+$ qubit ($B_0=12$\,mT with $\left|\uparrow\right>=\left|F=1,m_F=1\right>$ and $\left|\downarrow\right>=\left|F=2,m_F=0\right>$~\cite{Langer2005a}), a carrier $\pi$ pulse can be achieved in 1\,$\mu$s with $\tilde I$=15\,mA. For multi-qubit gates, an oscillating current $I(t)$ through the central conductor (a) produces a field $B_z(t)$ at the ion. Two currents $-2.5\cdot I(t)$ in conductors (c) and (d) produce a field $-B_z(t)$ at the ion, thereby nulling the field while providing field gradients $\partial B_x(t) / \partial z=\partial B_z(t) / \partial x = 2.5\cdot10^{-7} \cdot I(t)/d_0^2$\,T/m for motional-state excitation and multi-qubit gates. As an example, for the $\sigma_\varphi \sigma_\varphi$ gate consider the first-order field-insensitive $^9$Be$^+$ qubit (see above) and $d_0=30\,\mu$m. The gate implemented on a radial $z$ ``rocking'' or COM (center-of-mass) mode with $\omega_j=2\pi\cdot5$\,MHz could be realized in 20\,$\mu$s with $\tilde I = 1.7$\,A. For comparison, similar cw currents have been achieved in neutral atom magnetic chip traps of similar size. For the equivalent $\sigma_z\sigma_z$ gate, consider the $^9$Be$^+$ states $\left|\uparrow\right>=\left|F=2,m_F=2\right>$ and $\left|\downarrow\right>=\left|F=2,m_F=0\right>$, which can be reached from the field-independent qubit manifold with one microwave pulse, but have different magnetic forces, as required for the $\sigma_z\sigma_z$ gate. For a radial $x$ ``rocking'' or COM mode, the gate can be realized in the same time with $\tilde I = 1.3$\,A. 

\begin{figure}[tb]
  \centering
  \includegraphics[width=\columnwidth]{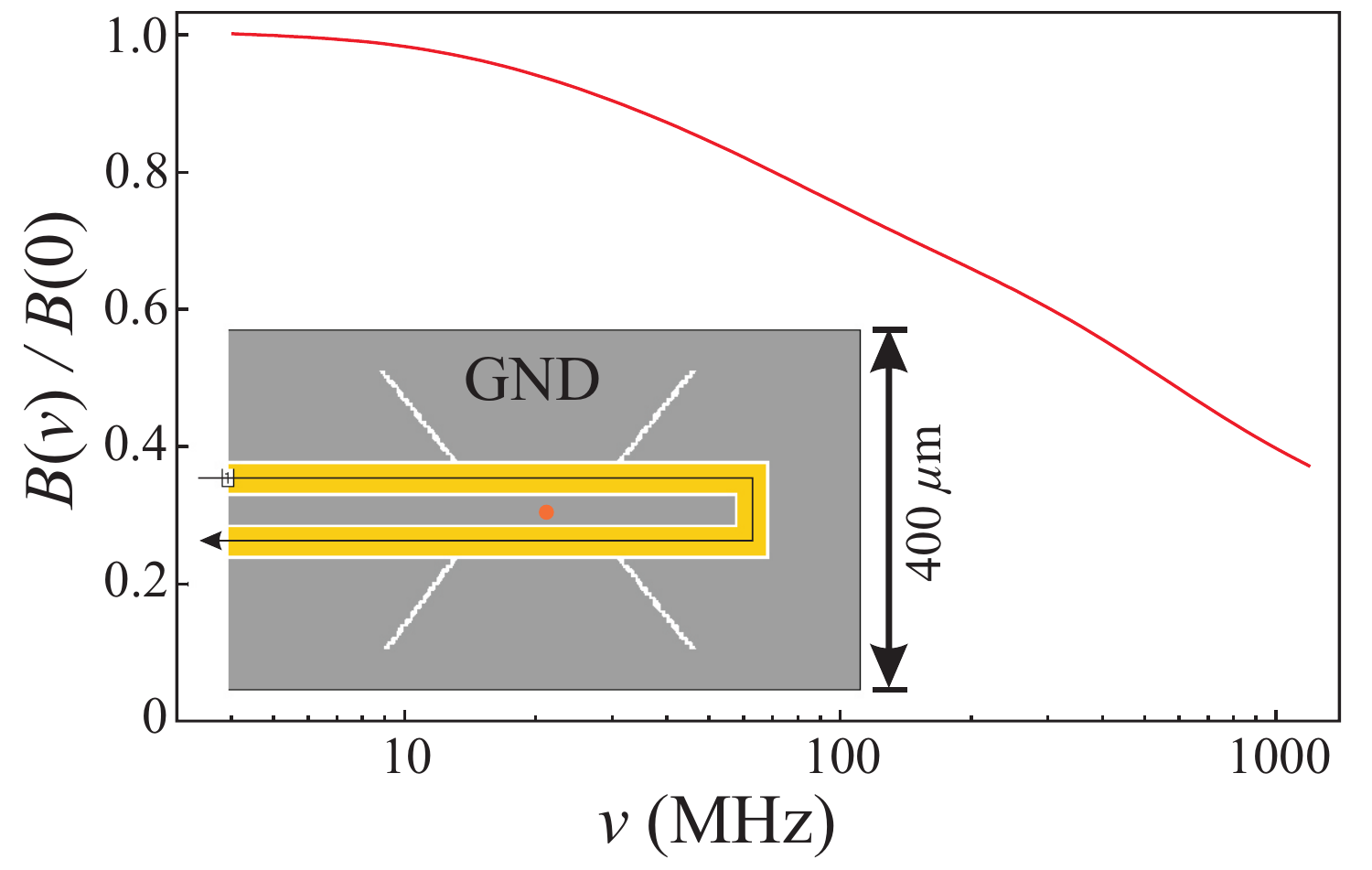}
  \caption{({\bf color online}) Simulation of surface current in an example trap structure, viewed from above. The figure shows the frequency dependence of the field 30\,$\mu$m above the center of the trap for the U-shaped current loop shown in the inset.}
\label{fig:rf_simulation.pdf}
\end{figure}

Practical implementation of these gates must account for a number of possible complications. Induced currents in neighboring trap electrodes will alter the magnetic fields and their gradients at the position of the ion. To estimate these effects as a function of frequency, we have performed finite-element numerical simulations of current flow in sample trap structures. The inset in Fig.~\ref{fig:rf_simulation} shows a U-shaped current loop, a geometry useful for single-qubit rotations. The center electrode and any surrounding trap electrodes are held at rf ground. The diagonal cuts in the outer plane serve to mimic the effect of the static potential control electrodes. Since the impedance of the circuit (ohmic vs.\ inductive) can change substantially over the considered frequency range, we have assumed that a fixed input power is dissipated in 6\,$\mu$m thick gold conductors ($\rho=2.44\cdot 10^{-8}\,\Omega$m) at each frequency. Fig.~\ref{fig:rf_simulation} shows the resulting magnetic field at a position $d_0=30\,\mu$m above the surface, normalized to the dc result. For frequencies up to about 20\,MHz (operating regime of a typical $\sigma_z\sigma_z$ gate), the resulting magnetic field is close to the dc value. 

In a large-scale device with multiple zones, it will be desirable to avoid ``crosstalk'' between zones. By adding more alternating currents to the geometry, it may be possible to achieve field decay with a desired polynomial order. Alternatively, compensating currents could be applied in neighboring trap zones to cancel crosstalk. For gate durations longer than a motional period, two-qubit gates will be effective only if the motional frequency is close to the gate-drive frequency. Therefore effects on ions in neighboring zones can be suppressed by confining them with different motional frequencies. 

Moreover, the currents applied to the electrode structures will likely be accompanied by oscillating electric potentials. This is of particular importance for the $\sigma_z\sigma_z$ gate on a COM mode, where the resulting oscillating electric field couples directly to the motional mode. For the $\sigma_z\sigma_z$ example discussed above, the (state-dependent) magnetic force on an ion in $\left|F=2,m_F=2\right>$ is equal to the (state-independent) electric force for an oscillating potential of $2.3\,\mu$V on electrode (a). Depending on the details of implementation and sources of electric fields, this effect can be compensated, because it yields only single-qubit phases, but these phases may be large and difficult to control. These effects are highly suppressed for $\sigma_z\sigma_z$ gates implemented on the rocking modes and the $\sigma_\varphi\sigma_\varphi$ gates.

Due to the strong field gradients desired for entangling gates, there will be significant oscillating fields if the ion is displaced from the field null. To avoid off-resonant carrier excitations and ac-Zeeman shifts, it may be necessary to make the current through conductors (c) and (d) in Fig.~\ref{fig:sample_geometry} adjustable, or to include additional `B-field compensation' wires to null the oscillating magnetic field, in analogy to micromotion compensation in ion traps~\cite{Berkeland1998a}. To estimate the effect of imperfect cancellation, we assume that the ions are located 200\,nm away from the field null along $z$. For either gate, the oscillating field will cause single-qubit phases no larger than 43\,mrad. These phases can be nulled by a suitable spin-echo sequence~\cite{Rowe2002a,Leibfried2003a} and/or (for the $\sigma_\varphi\sigma_\varphi$ gate) by a smooth overall global pulse envelope, because they commute with the gate propagator. Note that in the corresponding $\sigma_\varphi\sigma_\varphi$ laser gate, these off-resonant carrier excitations can also be avoided by placing the ions at the node of a standing-wave laser beam. However, due to the difficulty of locating the ion at a node, this gate has so far been realized only with running wave beams, where off-resonant carrier excitations proportional to $\sigma_{\pi/2-\varphi}$ occur. These do not commute with the gate propagator and therefore represent an important gate error~\cite{Sorensen2000a} that is absent in the magnetic gate.

An important benefit of magnetic-field gates is their potential insensitivity to motional-state initialization. In laser-mediated multi-qubit gates, it is usually necessary to initialize the ions very close to the motional ground state to avoid nonlinearity in the motional excitation. The extent of the motional ground-state wavefunction is on the order of 10\,nm, typically just one order of magnitude smaller than the spatial structure of the laser beams, given by $\lambda/(2\pi)$. Therefore, the ions will experience different values of $\Omega_{j,n}^x$ and $\Omega_{j,n}^z$ due to higher-order terms in the driving fields when they occupy different excited motional states. For the magnetic field gradient gates discussed here, the gate fidelity is insensitive to the initial motional state up to the point where the modes become anharmonic, as long as the field gradient is constant over the ion trajectory (as assumed in Eq.~(\ref{eq:hamiltonian})). Higher-order terms in the Rabi rates will be suppressed by a factor of $\approx({\tilde q}_0^j/d_0)^2$ compared to the linear terms. This should allow for gate operation on ions that are only Doppler-cooled, removing the need for near-ground-state cooling. Similarly, if ion motion is significantly excited during transfer in a multizone array, sympathetic Doppler cooling should be sufficient for high-fidelity gates. On the other hand, if ground-state cooling is desired, it can be implemented with the use of red-sideband transitions driven by oscillating magnetic field gradients and optical pumping. This motional-state insensitivity might also permit gate durations smaller than a motional period~\cite{GarciaRipoll2003a} because much larger trajectories in phase space can be tolerated.

There are additional important practical differences between the $\sigma_\varphi\sigma_\varphi$ gate and the $\sigma_z\sigma_z$ gate. The latter operates at a frequency close to the relatively low trap frequency, which will facilitate the generation of the required rf currents in a conductor structure. Also, the $\sigma_z\sigma_z$ gate can be readily applied to ``optical'' ion qubits where the two qubit states are, for example, S and D states as in Ca$^+$. The only requirement is that the two qubit states have different magnetic moments. The $\sigma_\varphi\sigma_\varphi$ gate, on the other hand, has the advantage that it can be applied directly to first-order magnetic field insensitive qubits and avoids the electric field excitation noted above. However, it would typically operate at a higher frequency of 1 to 15\,GHz, corresponding to hyperfine splittings, which makes it technically more difficult to implement. Exceptions include qubits based on the low-field Zeeman states of a $^2S_{1/2}$ level in an ion without nuclear spin.

In summary, we have discussed the implementation of universal gates for trapped-ion QIP with oscillating magnetic field configurations compatible with surface-electrode ion traps. These gates are free of spontaneous-emission decoherence and can be relatively insensitive to motional-state initialization, potentially eliminating the need for near-ground-state cooling. This could enable trapped-ion QIP with the use of rf/microwave fields and relatively low-power optical pumping, Doppler cooling and detection laser beams. The absence of phase, intensity, and beam-pointing instabilities that typically accompany laser-induced gates and the high degree of control possible with rf and microwave fields can be an important advantage in increasing the fidelity and reproducibility of gate operations. Given current trap technology, gate durations comparable to those of laser-mediated gates are achievable, with favorable scaling as trap dimensions become smaller. Recent observations of the suppression of anomalous motional heating at low temperatures~\cite{Deslauriers2006a,Labaziewicz2008a} may lead to much smaller dimensions (also not restricted by the diffraction limit on laser beams) and correspondingly higher gate speeds. 

We thank M. J. Biercuk and J. P. Home for helpful comments on the manuscript. We acknowledge funding by IARPA, ONR and NIST. This paper is a contribution of NIST, not subject to U. S. copyright.

\end{document}